\documentclass[12pt,letterpaper]{article}
\usepackage[letterpaper,ignoreall,top=2.65cm,bottom=2.7cm,left=2.65cm,right=2.65cm,foot=1cm,head=1.5cm]{geometry}
\usepackage{setspace}
\usepackage{xcolor}

\usepackage{times}
\usepackage{amsmath}
\usepackage{amssymb}
\usepackage{bm}
\usepackage{here}
\usepackage{graphicx}
\usepackage{framed}
\definecolor{shadecolor}{rgb}{0.9,0.9,0.9}
\usepackage[utf8x]{inputenc}

\usepackage{rsc}
\usepackage[below]{placeins}

\newlength{\figurewidth}
\begin{document}
\date{\today}

\begin{center}
\Large\textbf{The Microscopic Diamond Anvil Cell: Stabilization of Superhard, Superconducting Carbon Allotropes at Ambient Pressure}
\end{center}
\begin{center}
\textit{Xiaoyu Wang$^a$, Davide M Proserpio$^b$, Corey Oses$^{c,d}$, Cormac Toher$^{d,e,f}$, Stefano Curtarolo$^{c,d}$, and Eva Zurek$^{a,*}$} \\[1ex]
$^a$Department of Chemistry, \\ State University of New York at Buffalo, Buffalo, NY 14260-3000, USA \\[1ex]
$^b$Dipartimento di Chimica, \\ Universita' degli Studi di Milano, Via Golgi, 19 - 20133 Milano, Italy \\[1ex]
$^c$Department of Mechanical Engineering and Materials Science, \\ Duke University, Durham, NC 27708, USA \\[1ex]
$^d$Center for Autonomous Materials Design, \\
Duke University, Durham, NC 27708, USA \\[1ex]
$^e$Department of Materials Science and Engineering, \\
University of Texas at Dallas, Richardson, TX 75080, USA\\[1ex]
$^f$Department of Chemistry and Biochemistry, \\
University of Texas at Dallas, Richardson, TX 75080, USA\\[1ex]
E-mail: ezurek@buffalo.edu
\end{center}

A metallic covalently bonded carbon allotrope is predicted via first principles calculations. It is composed of an $sp^3$ carbon framework that acts as a diamond anvil cell by constraining the distance between parallel cis-polyacetylene chains. The distance between these $sp^2$ carbon atoms renders the phase metallic, and yields two well-nested nearly parallel bands that span the Fermi level. Calculations show that this phase is a conventional superconductor, with the motions of the $sp^2$ carbons being key contributors to the electron phonon coupling. The $sp^3$ carbon atoms impart superior mechanical properties, with a predicted Vickers hardness of 48~GPa. This phase, metastable at ambient conditions, could be made via cold compression of graphite to 40~GPa. A family of multifunctional materials with tunable superconducting and mechanical properties could be derived from this phase by varying the $sp^2$ versus $sp^3$ carbon content and by doping.

\newpage

Metallic covalently-bonded materials are candidates for conventional, or phonon-mediated, superconductivity \cite{Blase2009}. Vibrations associated with the metallic covalent bonds, such as the B-B $\sigma$ bonds in MgB$_2$ \cite{An:2001,Kortus:2001}, the C-C $sp^3$ bonds in boron doped diamond \cite{Blase:2004,Boeri:2004}, and the weak multi-centered H-H bonds in the hydrogenic clathrate cages of compressed superhydrides \cite{Zurek:2018m} are characterized by a large electron phonon coupling (EPC). The light mass of the constituent elements and the large density of states (DOS) at the Fermi level ($E_F$) is key to achieving a high superconducting critical temperature, $T_c$.  One way metallic covalent materials can be made is via the formation of unusual bonding environments induced by the high pressures present within diamond anvil cells. Carbon is particularly attractive since its strong bonds result in large kinetic barriers, important for quenching metastable materials to atmospheric pressures, and enhanced mechanical properties such as high density, superior hardness and large bulk modulus. Herein, density functional theory (DFT) calculations are performed to propose a form of carbon that is superconducting and superhard, and could be synthesized under mild pressures. Because it is characterized by an $sp^3$ framework that behaves as a microscopic diamond anvil cell by constraining the distance between $sp^2$ carbons, key for its metallicity, we call it DAC-carbon. Modifications of this structure could lead to a family of superconducting, superhard, multifunctional materials with tunable properties.
\begin{figure}[h!]
    \centering
    \includegraphics[width=0.55\columnwidth]{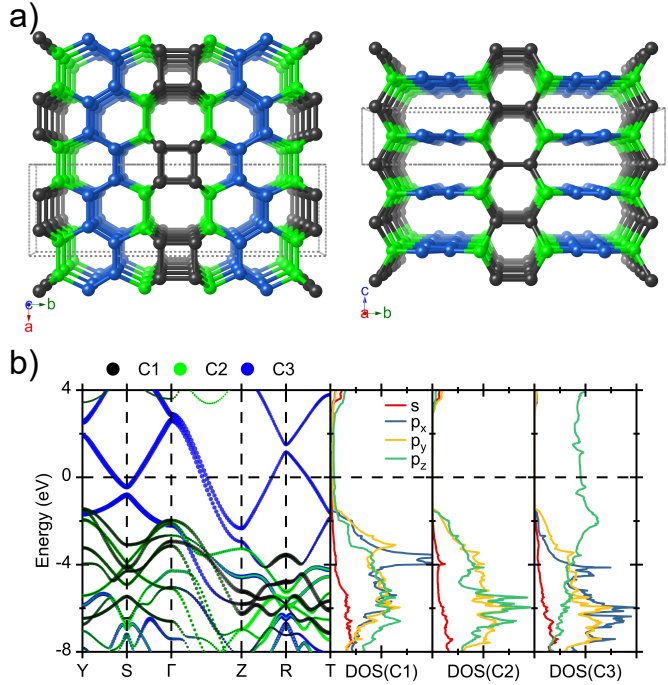}
    \caption{(a) Optimized geometry of DAC-carbon with top view and side view. Its standard primitive cell contains 12 carbon atoms (8 $sp^3$ and 4 $sp^2$) and possesses the $Cmmm$ spacegroup. Black balls are $sp^3$ carbon that are bonded only to other $sp^3$ carbons, green balls are $sp^3$ carbons bonded to both $sp^3$ and $sp^2$ carbons, and blue balls are $sp^2$ carbons; dashed lines denote the conventional cell. (b)  Band structure along the $Y (0.5, 0.5, 0) \rightarrow S (0, 0.5, 0) \rightarrow \Gamma (0, 0, 0) \rightarrow Z (0, 0, 0.5) \rightarrow R (0, 0.5, 0.5) \rightarrow T (0.5, 0.5, 0.5)$ high symmetry lines, and orbital projected density of states (DOS, in states eV$^{-1}$\AA{}$^{-3}$) for DAC-carbon. The thickness of the lines in the band structure denotes the contribution from the listed atom types.}
    \label{fig:structure}
\end{figure}

Previously, we predicted low-energy superhard carbon allotropes using a multi-objective evolutionary algorithm that employed both the DFT energy, and Vickers hardness ($H_\text{v,Teter}$) estimated using shear moduli obtained via a machine learning (ML) model trained on the AFLOW database.~\cite{Avery2019}  Forty-three novel superhard phases were found, and the topological properties of their carbon frameworks were analyzed. However, their electronic structures, bonding peculiarities, and propensity for superconductivity were not discussed. 

Though most of the novel carbon allotropes were insulators with large gaps between the conduction and valence band, some were semiconductors, and two were metallic.  DAC-carbon, referred to as $Cmmm$-12b in Ref.\ \cite{Avery2019}, is one of these. Its primitive cell can be constructed by inserting  $sp^2$ carbon atoms in an all-cis-polyacetylene chain into the $sp^3$ framework of the quasi-lonsdaleite structure R$_2$L$_2$~\cite{Baughman1997_Carbon}, commonly referred to as $Z$-carbon, and listed in the SACADA database \cite{Hoffmann2016_SACADA} with the topology \textbf{sie} (Figure \ref{fig:structure}). At zero pressure DAC-carbon was  230~meV/atom (5.37~kcal/mol) less stable than diamond (within the PBE functional), and its enthalpy fell below that of graphite above 40~GPa. Phonon calculations confirmed this phase was dynamically stable from 0-5~GPa at 0~K, and molecular dynamics (MD) simulations at 100, 200, 300 and 400~K on the 20 and 40~GPa ground state geometries illustrated they were kinetically stable.  These results suggest that DAC-carbon could potentially be synthesized under pressure, similar to the $M$-carbon phase (\textbf{cbn} in SACADA \cite{Hoffmann2016_SACADA}) formed upon the cold compression of graphite \cite{mao2003bonding}. Moreover, our MD simulations on the zero pressure geometry showed that the $sp^2$ carbon atoms from adjacent polyacetylene layers only begin to interact at 1600~K, suggesting the kinetic barriers to decomposition are large. Using the ML (DFT) calculated shear modulus we found $H_\text{v,Teter}$~=~45 (48)~GPa, as compared to 72~GPa for R$_2$L$_2$~\cite{Avery2019}. Insertion of the $sp^2$ atoms into R$_2$L$_2$ dramatically decreases its hardness, however thanks to its $sp^3$ framework DAC-carbon still falls above the superhard threshold.

The $sp^2$ chains in DAC-carbon propagate along the $a$-axis and are stacked along the $c$-axis. Their interlayer distance of 2.57~\AA{}, dictated by the rigid framework of the $sp^3$ carbons, is considerably smaller than within the cis-polyacetylene crystal, with measured inter-chain distances of 4.4~\r{A} \cite{Baughman1978}. This geometrical feature of DAC-carbon is reminiscent of the infinite polyene chains imagined by Hoffmann et al., whose interchain distance of 2.50~\AA{} induced the metallicity in this hypothetical $sp^2$ carbon allotrope~\cite{Hoffmann1983}. Hoffmann's work inspired the theoretical prediction of other 3D forms of carbon whose metallicity was induced by the steric confinement of the $sp^2$ carbon atoms~\cite{Bucknum1994_Glitter}. Many of the predicted phases possessed a high hardness due to their large $sp^3$ ratio~\cite{wu2017,liu2020,liu2018}, and the $T_c$ of two that were not superhard was predicted to be 5 and 14~K,~\cite{Hu2015} but the mechanism of superconductivity was not analyzed. 

\begin{figure}[b!]
    \centering
    \includegraphics[width=1\columnwidth]{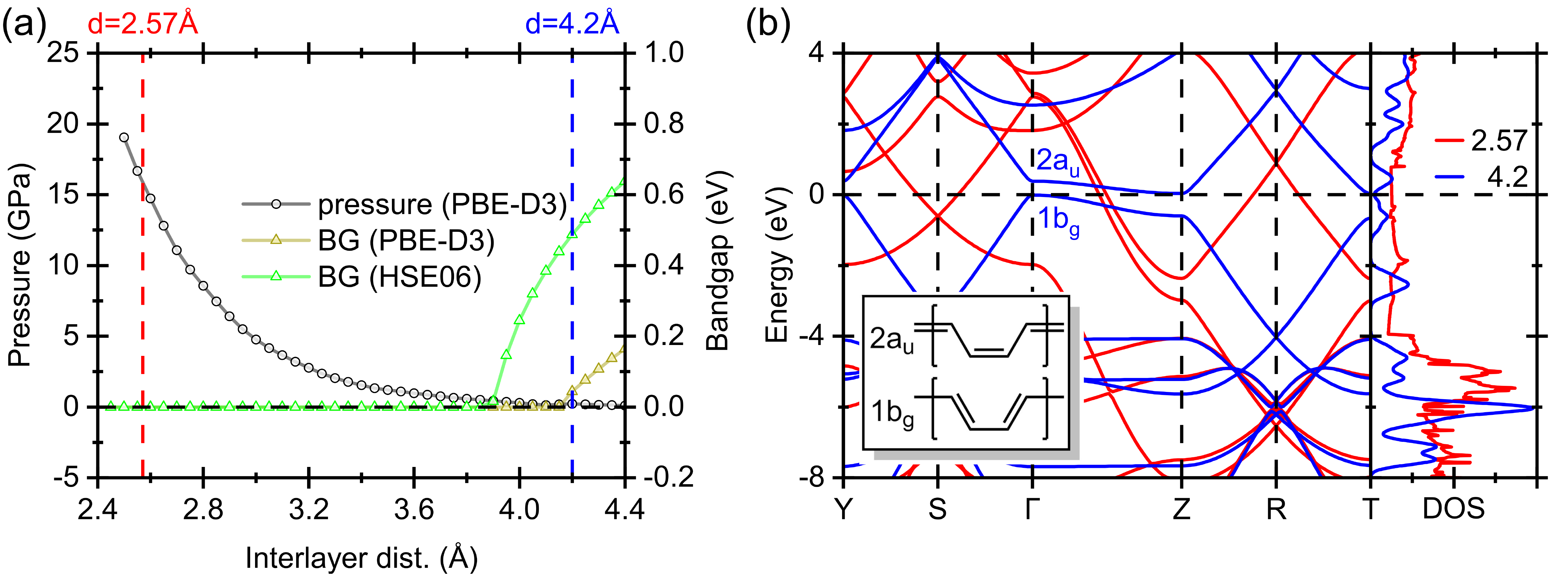}
    \caption{(a) Band gap of an ensemble of cis-polyacetylene chains for the given interlayer distances as calculated with the non-hybrid PBE-D3 (brown triangles) and hybrid HSE-06 (green triangles) functionals. The pressure at these interlayer distances as obtained with PBE-D3 is also provided (black circles). (b) Band structure as computed with the PBE functional for the cis-polyacetylene chain for interlayer distances of 4.20~\AA{} (blue) and 2.57~\AA{} (red). Since the interlayer spacing affects the $c$ lattice constant the length along the $\Gamma-Z$ high-symmetry lines for the non-interacting chains has been scaled to match those whose distance is constrained to be the same as in DAC-carbon.}
    \label{fig:model}
\end{figure}

The band structure and DOS plots of DAC-carbon (Figure \ref{fig:structure}(b)) clearly illustrates that their metallicity stems from the $p_z$ orbitals of the $sp^2$ carbons. To explore this further we built a model where the $sp^3$ carbons of DAC-carbon, whose bond distances were nearly equal to those within diamond (1.57 vs.\ 1.55~\AA{}), were removed from the cell and the dangling bonds were saturated by hydrogens. The resulting layered cis-polyacetylene chain possessed a repeating C$_4$H$_4$ unit where the C-C bonds measured 1.39 and 1.40~\AA{} (c.f.\ 1.37~\AA{} calculated for cis-polyacetylene), as in the relaxed DAC-carbon structure. Varying the interlayer distance, we computed the bandgap and estimated the internal pressure caused by the confinement from the negative of the change in energy versus volume, $P=-dE/dV$, as obtained numerically via the central difference method. Figure \ref{fig:model}(a) illustrates that within the PBE-D3 (HSE-06) functionals the band gap closed when the interlayer distance was 4.15 (3.85)~\AA{} corresponding to a PBE-D3 pressure of 0.19 (0.50)~GPa. At the distance found in the optimized DAC-carbon lattice, the model system remained metallic and the internal pressure was calculated to be 15.9~GPa. \emph{Thus, the lattice of $sp^3$ carbons comprising DAC-carbon can be thought of as a microscopic diamond anvil that exerts pressure on the cis-polyacetylene chain, thereby inducing metallicity.}

To better understand the origin of the insulator-to-metal transition, the PBE band structure of the cis-polyacetylene chain (Figure \ref{fig:model}(b)) was analyzed. When the interaction between neighboring layers is small the highest occupied (lowest unoccupied) crystal orbital at the Zone center corresponds to the $1b_g$ ($2a_u$) symmetry linear combination of $p_z$ orbitals that are $\pi$ bonding (anti-bonding) along the shorter, and $\pi$ anti-bonding (bonding) along the longer C-C distance, as illustrated schematically in the inset. At the $\Gamma$-point they are $p_z$ $\sigma$ anti-bonding with the next layer, whereas at the $Z$-point the $p_z$ $\sigma$ interaction is favorable. This interaction, insignificant at large distances due to the negligible orbital overlap, becomes increasingly important as the interlayer distance decreases. When the distance is the same as in DAC-carbon, the bands are pushed high above $E_F$ at the Zone center, and they run down to the $Z$-point nearly parallel to one another, as first proposed by Hoffmann for an all-$sp^2$ carbon analogue of the ThSi$_2$ structure~\cite{Hoffmann1983}. Near $E_F$ the band structure we calculate for the squeezed cis-polyacetylene chain model is in strikingly good agreement with the bands obtained for DAC-carbon (cf.\ Figure \ref{fig:model}(b) and Figure \ref{fig:structure}(b)). Though the $p_z$ $\sigma$ bonding within DAC-carbon is weak (the integrated Crystal Orbital Hamilton Population for nearest neighbors is -0.06~eV/bond, cf.\ -9.5~eV/bond for the bonds in diamond), it is key for the metallicity of this phase.

This set of steep parallel bands separated by $\sim$0.48~eV suggests a Fermi surface that is well nested. Could DAC-carbon be a covalently bonded conventional superconductor? To answer this question we calculated the phonon band structure, Eliashberg spectral function, the EPC parameter ($\lambda=0.37$), and logarithmic average of the phonon frequencies ($\omega_\text{ln}=670$~K) for this phase. Within the Allen-Dynes modified McMillan equation, and using a renormalized Coulomb repulsion parameter characteristic of boron doped diamond, $\mu^*=0.1$, $T_c$ was estimated to be 1.6~K. This value is strikingly close to the only known superconducting form of pure carbon, magic angle twisted bilayer graphene ($T_c=$~1.7~K), whose superconductivity is thought to be a result of strong electron correlations~\cite{Cao2018}, and somewhat lower than that of boron doped diamond ($T_c=$~4~K for a doping level of 2.5\%)~\cite{Ekimov2004}. The $\lambda$ of DAC-carbon is similar to estimates for diamond doped with 1.85\% boron \cite{Blase:2004}, even though its DOS at $E_F$ is about a factor of five smaller. 

\begin{figure}[!b]
    \centering
    \includegraphics[width=1\columnwidth]{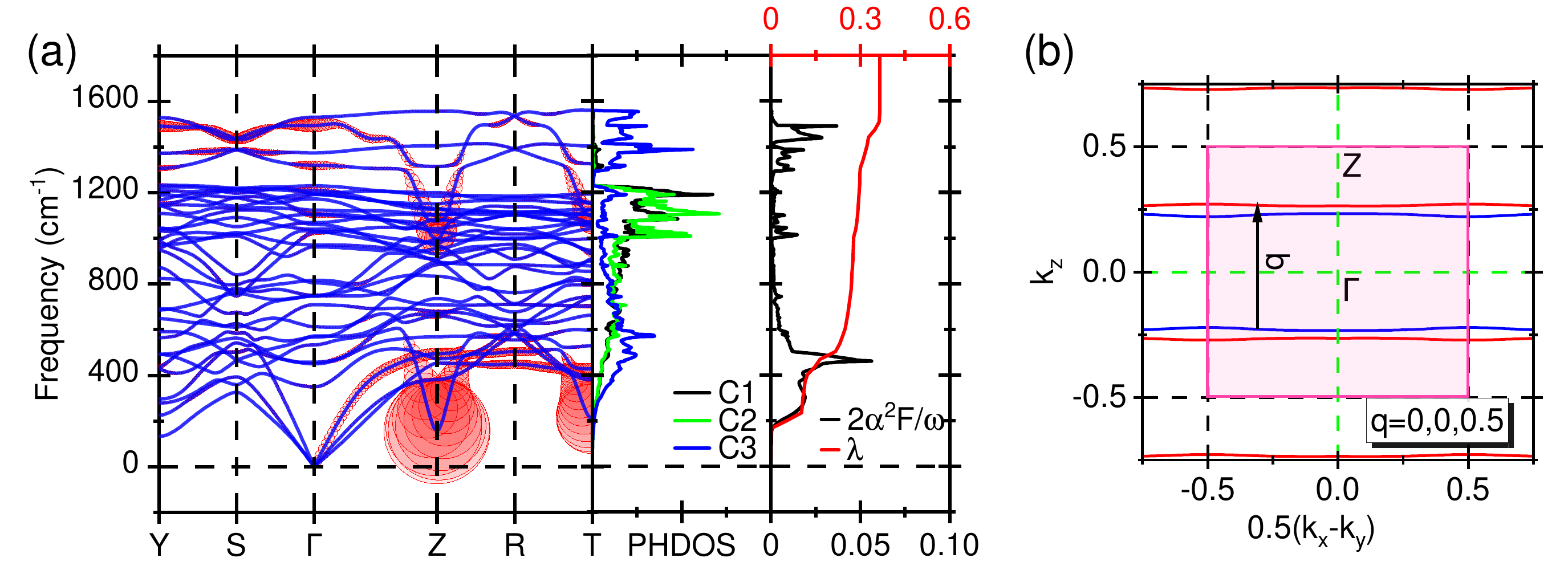}
    \caption{(a) Phonon band structure, atom projected phonon density of states (PHDOS), Eliashberg spectral function, in the form of $\frac{2\alpha ^2F(\omega)}{\omega}$, and the electron phonon integral, $\lambda(\omega)$, for DAC-carbon. Red circles indicate the electron-phonon coupling constant, $\lambda_{\textbf{q}\nu}$, at mode $\nu$ and wavevector $\textbf{q}$, and their radii is proportional to the strength. (b) Isocontour of eigenvalues for the two parallel bands $1b_g$ (blue) and $2a_u$ (red) at the Fermi level. The pink shaded region represents the first Brillouin zone. Black vector indicates the phonon vector $\textbf{q}=(0,0,0.5)$ by which the two bands are nested.}
    \label{fig:phonons}
\end{figure}

To analyze the nature of the pairing mechanism, we plotted the phonon band structure decorated by the EPC line-widths, whose thickness is proportional to the coupling strength (Figure \ref{fig:phonons}(a)). A soft mode with a frequency of 156~cm$^{-1}$ at the $Z$-point had the largest contribution, 26\%, to the total $\lambda$. About 20\% of $\lambda$ was due to the four highest frequency bands, which are associated with the in-plane stretching modes of the $sp^2$ carbons. These bands are relatively flat, but soften significantly around the $Z$-point. Careful inspection of the Fermi surface plots showed that the two parallel bands crossing $E_F$ along the $\Gamma\rightarrow Z$ line are strongly nested, so that an electron travelling on one of the surfaces can absorb a phonon with wavevector $\textbf{q}=(0,0,0.5)$ and be scattered on the other surface resulting in a large EPC ((Figure \ref{fig:phonons}(b)). Visualization of the vibration associated with the 156~cm$^{-1}$ mode showed that it corresponded to the rotation of a pair of $sp^2$ carbon atoms in the $cb$ plane, with dimers in neighboring layers rotating in opposite directions, while the $sp^3$ carbons remain stationary. This motion modified the distance between carbon atoms comprising neighboring polyacetylene chains, and the $p_z$ $\sigma$ overlap between them. Moving the atoms along this eigenvector, and optimizing to the nearest local minimum, resulted in a dynamically stable fully $sp^3$ hybridized structure with $Cmcm$ symmetry, which was superhard ($H_\text{v,Teter}^\text{ML}=$~68~GPa), insulating, and $\sim$11~meV/atom less stable than DAC-carbon. This structure has not yet been reported in the literature, however it shares the same topology as a hypothetical zeolite \cite{Blatov2020}. At the $T$-point these same sets of bands soften and possess a substantial, albeit slightly smaller, $\lambda_{\textbf{q}\nu}$.  Similar to what was found at the $Z$-point, this phonon wavevector also led to the nesting of the Fermi surfaces arising from these two parallel sets of bands, however in a higher-order Brillouin zone. Visualization of the softest mode at 210~cm$^{-1}$ revealed that it was associated with $p_z$ $\sigma$ overlap between $sp^2$ hybridized carbons in the polyacetylene chains, and following this mode resulted in the formation of a novel all-$sp^3$ $Ibam$ symmetry carbon allotrope that is 6~meV/atom less stable than DAC-carbon.

Owing to the relatively heavy mass of carbon, the highest vibrational frequency in DAC-carbon is $\sim$1600~cm$^{-1}$, and the $\omega_\text{ln}$ is relatively low.  Because only the $p_z$ orbitals of the $sp^2$ carbons contribute to the metallicity, the DOS at $E_F$ is also low. Both of these $T_c$ descriptors could be increased via boron doping, and $\omega_\text{ln}$ could be improved by inserting H$_2$ into the voids within the R$_2$L$_2$ lattice. Since the weak $p_z$ $\sigma$ interaction, which is dependent on the interlayer distance, is key for the EPC, the $T_c$ is likely pressure dependent. Indeed, our calculations show  $T_c$ increases to 8.3~K at 5~GPa.  The R$_2$L$_2$ strips could be thickened thereby hardening the allotrope, but widening the $sp^2$ chains would soften the material and increase the number of states participating in the EPC mechanism. Finally, different $sp^3$ frameworks that comprise the microscopic diamond anvil can be chosen. We dream some of these multifunctional allotropes, where the carbon framework acts as a diamond anvil cell, can be experimentally realized.

\section*{Acknowledgements}
E.Z.\ and X.W.\ acknowledge the NSF (DMR-2119065), E.Z., C.T., C.O., and S.C. acknowledge the DOD-ONR (N00014-16-1-2583), and C.T., C.O., and S.C. acknowledge the DOD-ONR MURI program (N00014-15-1-2863) for financial support. The Center for Computational Research (CCR)  (http://hdl.handle.net/10477/79221) at SUNY Buffalo is acknowledged for computational support.

\providecommand*{\mcitethebibliography}{\thebibliography}
\csname @ifundefined\endcsname{endmcitethebibliography}
{\let\endmcitethebibliography\endthebibliography}{}

\clearpage
\newpage

\noindent\textbf{Keywords:} carbon allotropes, density functional calculations, electronic structure, superconductors, superhard materials \\[2ex]

\noindent\textbf{Table of Contents Graphic:} \\
\begin{figure*}[h!]
\centering\includegraphics[width=5.5cm]{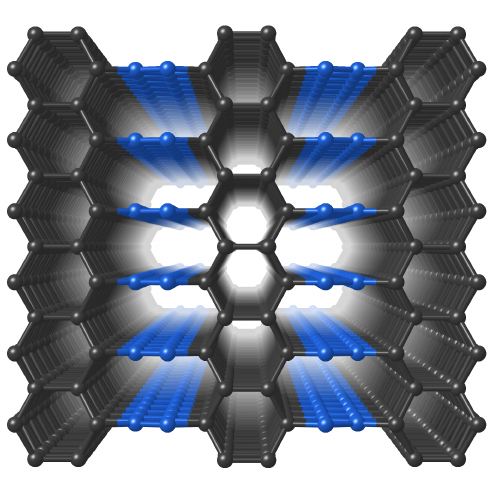}
\end{figure*} \\[2ex]

\noindent\textbf{Table of Contents Text:}\\

\noindent\textbf{DAC-Carbon:} The superhard $sp^3$ framework serves as a microscopic diamond anvil cell, in
which the $sp^2$ chains are compressed to establish the metallicity and superconductivity.

\end{document}